\documentclass[conference]{IEEEtran}
\IEEEoverridecommandlockouts

\usepackage{cite}
\usepackage{url}
\usepackage{bbm}
\usepackage{amsmath,amsfonts} 
\usepackage{float}
\usepackage{graphicx}
\usepackage{algorithm}
\usepackage{algpseudocode}
\usepackage{siunitx}
\usepackage[labelfont=bf]{caption}
\usepackage{subcaption}
\usepackage{tikz}
\usepackage{pgfplots}
\usepackage{epstopdf}
\usepackage{multirow}
\usepackage[font=scriptsize]{caption}
\usepackage{soul}
\usepackage{color, colortbl}
\usepackage{array}
\usepackage[T1]{fontenc}
\usepackage[utf8]{inputenc}
\usepackage{mathptmx}
\usepackage{listings}
\usepackage{adjustbox}
\usepackage{makecell}
\usepackage{censor}

\usepackage{enumitem}
\usepackage{hyperref}
\usepackage{pifont}

\def\BibTeX{{\rm B\kern-.05em{\sc i\kern-.025em b}\kern-.08em
    T\kern-.1667em\lower.7ex\hbox{E}\kern-.125emX}}
    
\makeatletter
\def\endthebibliography{%
  \def\@noitemerr{\@latex@warning{Empty `thebibliography' environment}}%
  \endlist
}
\makeatother

\begin{document}
\StopCensoring

\title{Malicious Lateral Movement in 5G Core With Network Slicing And Its Detection}


\author{\IEEEauthorblockN{Ayush Kumar and Vrizlynn L.L. Thing} \\
\IEEEauthorblockA{Cyber Security Strategic Technology Centre \\
Singapore Technologies Engineering \\
Email: ayush.kumar@u.nus.edu, vriz@ieee.org}
}
			


\maketitle

\begin{abstract}
5G networks are susceptible to cyber attacks due to reasons such as implementation issues and vulnerabilities in 3GPP standard specifications. In this work, we propose lateral movement strategies in a 5G Core (5GC) with network slicing enabled, as part of a larger attack campaign by well-resourced adversaries such as APT groups. Further, we present \textsf{5GLatte}, a system to detect such malicious lateral movement. \textsf{5GLatte} operates on a host-container access graph built using host/NF container logs collected from the 5GC. Paths inferred from the access graph are scored based on selected filtering criteria and subsequently presented as input to a threshold-based anomaly detection algorithm to reveal malicious lateral movement paths. We evaluate \textsf{5GLatte} on a dataset containing attack campaigns (based on MITRE ATT\&CK and FiGHT frameworks) launched in a 5G test environment which shows that compared to other lateral movement detectors based on state-of-the-art, it can achieve higher true positive rates with similar false positive rates.
\end{abstract}  


\section{Introduction}
\label{intro}
5G is the most recent standard for mobile communication and provides major enhancements compared to earlier cellular technologies. With 5G, high data transfer rates (up to 10 Gbps), very low latency (below one millisecond),  high-speed mobility (up to 500 km/h) and high connection densities (one million per square kilometer) are possible \cite{5g-netw-perf}. 5G introduces a new network architecture called \textit{network slicing} that allows for the creation of multiple virtualized and independent logical networks on the same physical network infrastructure where each network slice is tailored to meet the diverse requirements of a specific application. Since 5G networks are IP-based, it opens them to a host of cyber attacks which were not possible with previous generation mobile networks. Though 3GPP standards have mandated several security features for 5G networks (inter-operator security provided by security proxy servers, subscriber identifier encryption for privacy, mutual authentication for network and devices), implementation issues can introduce vulnerabilities in telecommunication networks. Sometimes the 3GPP standard specifications are themselves vulnerable to exploitation \cite{bookworm}. Several such vulnerabilities have been discovered by academic and industry researchers.

Cyber attacks targeting critical and complex infrastructure, such as 4G/5G networks, consist of multiple stages with \textit{lateral movement} being one of them, as outlined by the MITRE FiGHT framework. Lateral movement  (LM) is a technique used by cyber attackers to progressively move through a network in search of targeted key data and assets. After gaining initial access to an endpoint, such as through a phishing attack or malware infection, the attacker moves through multiple systems in the network until the end goal is reached. Attaining that objective involves gathering information about multiple systems and accounts, obtaining credentials, escalating privileges and ultimately gaining access to the identified payload.


In this work, we first present malicious LM strategies in 5G Cores (5GC) with network slicing. The main idea behind this attack is for an attacker with a foothold in the 5GC to pivot to a connected network slice or move across slices. This attack can enable cyber attackers to, (1) reach NFs which are not directly connected to the NFs they have access to and (2) establish foothold in critical network slices (e.g., e-health, autonomous mobility, industrial IoT) which they may not have initial access to. Such LM may be more applicable to private 5GC implemented by commercial vendors who might pay less attention to security than the 5GC deployed by large telecommunication networks, though they are not immune to cyber attacks as evidenced by the attack on Vodafone Portugal reported in February 2022 \cite{vodafone-attack}. Existing proprietary as well as open-source 5GC are known to have several security vulnerabilities \cite{trend-micro-5gc-vuln} and are therefore, prone to compromise. 

We have also proposed a system to detect such LM using host logs and NF container logs, \textsf{5GLatte}. A host-container access graph is built using the logs. Paths inferred from the access graph are scored based on selected filtering criteria and subsequently subjected to an anomaly detection algorithm, revealing LM which is most likely to be malicious. In the absence of any existing datasets containing LM in a 5GC, we build a 5GC testbed based on the OAI open-source implementation of the 3GPP standard and collect data from it both during normal operation and under attack. To simulate the attack scenario, we run realistic multi-stage attack campaigns (including LM as one of the stages) based on MITRE ATT\&CK and FiGHT frameworks. We evaluate \textsf{5GLatte} on this dataset and compare its performance with other lateral movement detectors based on state-of-the-art.

The main contributions of our work are as follows: 
\begin{itemize}
	\item We propose and study malicious lateral movement strategies in 5GC with network slicing.
	\item We design \textsf{5GLatte} to detect such lateral movement in 5GC using threshold-based anomaly detection.
	\item We evaluate \textsf{5GLatte}'s performance based an a realistic dataset collected from our 5GC test environment.
\end{itemize}
 

\section{Related Work}
\label{literature}

\subsection{Cyberattacks on 5G Network Slicing and Secure Slice Management}
\label{5gc-slice-attack}
A number of cyberattacks on 5G network slices have been proposed in literature. In \cite{dsm-attack}, Sathi et al. have proposed a distributed slice mobility (DSM) attack in 5G networks. If a UE is compromised by an attacker, it can initiate an inter-slice mobility request to migrate from origin slice (where UE is registered) to a destination slice. The migration can cause fluctuating traffic loads in destination slice and thus performance damage to on-going sessions of UEs registered to destination slice. Extrapolating the above example, if there are multiple compromised UEs, they might migrate back and forth between a target network slice and other random slices, remaining in the target slice for a certain time duration, causing a Distributed Denial-of-Service (DDoS) on the target slice. In \cite{shi-flood}, Shi et al. have introduced a flooding attack on 5G network slicing, where an adversary generates fake network slicing requests to consume the 5G RAN resources that would otherwise be available to real requests. The adversary observes the spectrum and builds a surrogate model on the network slicing algorithm through Reinforcement Learning that decides on how to craft fake requests to minimize the reward of real requests over time. 


A few works have also proposed secure network slice management approaches. \cite{kholidy} has included an autonomous secure network slicing system which ingests VNF logs from network slices, feeds them to a deep learning model whose output is used by a risk assessment model to calculate the risk of deploying a user task to a network slice based on which the policy decision for task deployment is taken. \cite{benzaid} has presented a security management framework which provides zero-touch security management by introducing intelligent closed-loops with different scopes and timescales, from the network functions to the inter-slice and end-to-end levels of the 5G architecture.   

\textbf{Difference from State-of-the-Art}: In \cite{dsm-attack,shi-flood}, the cyberattacker is assumed to have gained control of UEs (connected to a gNB base station) which make periodic inter-slice mobility requests, or generate fake network slice requests, or target network slices with DDoS attacks respectively. Our proposed attack instead assumes that the cyber attacker has compromised one or more NF containers (\cite{dark-reading,bleeping-comp}) in a 5G core network. \cite{kholidy,benzaid} propose secure network slice management approaches. The former work focuses on risk assessment of user task allocation to a network slice (no slice attack detection) while the latter work focuses on anomaly detection using deep learning models on slice resource usage and performance metrics (which may work for sustained attacks such DDoS). However, they are not designed to detect exploitation of vulnerabilities on NF containers used to orchestrate slices which forms an integral part of our proposed LM strategies.

\subsection{Lateral Movement Detection in Enterprise Networks} 
Since our work focuses on LM detection in 5G cores with network slicing and there being no prior work in that area, we discuss a few prominent works on LM detection in enterprise IT networks. Hopper \cite{hopper} targets LM detection by building a graph of login activity among machines in a network and identifying suspicious login sequences. The broader paths that each login belongs to are then identified and finally, an anomaly detection algorithm is applied to conservatively infer the set of login paths most likely to reflect LM. Latte \cite{latte} splits the LM detection problem into two parts: \textit{forensic analysis}, which detects inbound and outbound LM paths from a known compromised computer or account, and \textit{general detection}, which identifies previously unknown LM in a network by integrating a remote file execution detector with a rare path detector. Bohara et al. \cite{bohra-lmdet} have presented an attack-vector agnostic approach to detect LM by first building a host-communication graph, extracting features from it and then correlating the outputs of a C\&C activity detector using MAD-based anomaly detection and an LM activity detector using PCA and \textit{k}-means on the extracted features to find infected hosts. 

Instead of manually extracting LM-related features from enterprise network logs, Bowman et al. \cite{graph-raid} have proposed to use an unsupervised graph learning technique on a graphical data structure that models authentication activity in a given network to detect malicious authentication events. Their LM detection pipeline consists of looking up node embeddings for an authentication graph corresponding to new authentication events and subsequently predicting links with low probabilities in combination with threshold-based anomaly detection. EULER \cite{euler} takes it a step further by considering the network as a dynamic graph with temporal links between nodes. It combines a Graph Neural Network with a sequence encoding algorithm, both of which are model-agnostic, to predict and detect anomalous temporal links.

\textbf{Difference from State-of-the-Art}: Prior works on LM detection \cite{hopper,latte,bohra-lmdet,graph-raid,euler} have targeted enterprise IT networks whereas the focus of our work is 5G core networks. 5GC architecture has unique components such as NF containers and network slices which warrant a different approach for LM detection. Compared to \cite{latte}, our work does not use forensic analysis or general detection as those techniques are designed for Windows machines only, assume either a known compromised computer or remote file execution and use a different path scoring mechanism. \cite{bohra-lmdet} assumes the presence of C\&C activity unlike our work. The graph learning-based malicious link detectors in \cite{graph-raid} and \cite{euler} are suitable for large enterprise networks and are trained/evaluated on public datasets containing large number of authentication events (e.g., 2015 LANL Comprehensive Cyber Security Events dataset \cite{lanl-dataset}) unlike our work which targets LM in a 5GC which has smaller number of nodes compared to an enterprise network and further, there are no LANL-like datasets available for 5GCs.

\section{Background}
\subsection{5G Service-Based Architecture}
The mobile cellular network enables wireless connectivity for devices such as smartphones and tablets, referred to as User Equipment (UE).
It is made up of two primary subsystems: the Radio Access Network (RAN) and the Mobile Core. The RAN is responsible for managing and ensuring efficient usage of radio resources such as spectrum and meeting users' Quality of Service (QoS) requirements. It consists of several base stations (each covering a particular area), known as eNB (`e'volved Node B) in 4G and gNB (next `g'eneration Node B) in 5G. 
The Mobile Core is a collection of functions that provides several services including authenticating devices before allowing them to connect to the network, ensuring that the connectivity meets QoS requirements, providing IP connectivity for data-based services, tracking device mobility and tracking subscriber usage.

\begin{figure*}[t]
	\caption{5G Mobile Core Architecture}
	\label{5gc-arch}
	\centering
	\includegraphics[scale=0.35]{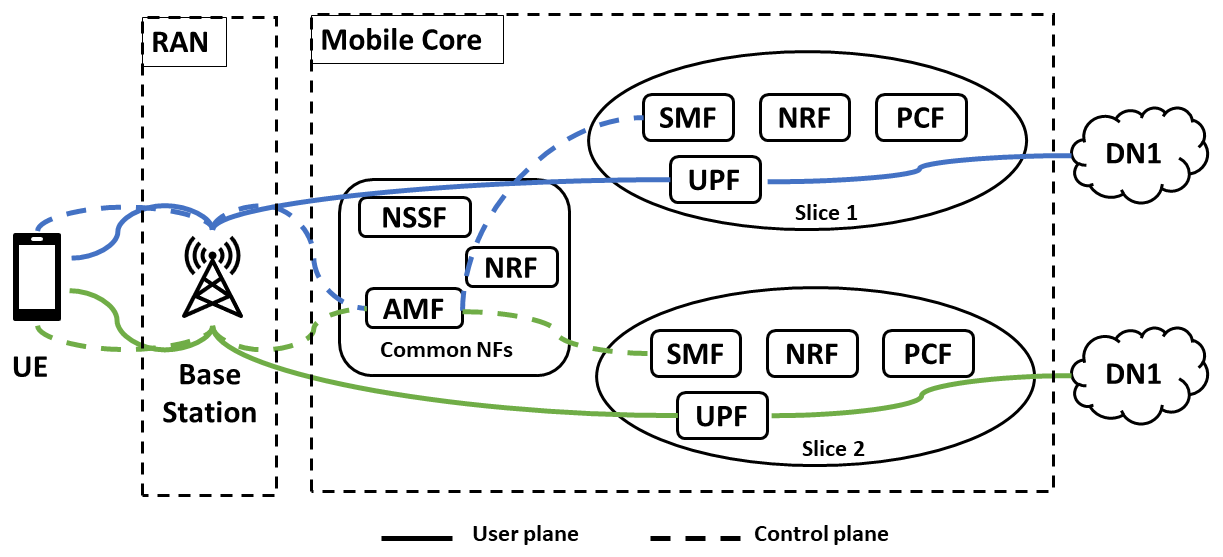}
\end{figure*}

The 5G Mobile Core (5GC) adopts a microservice-like architecture (shown in Fig. \ref{5gc-arch}), with a set of functional blocks, each of which is called a Network Function (NF). The NFs are divided into three groups, with two groups running in the Control Plane (CP) and one group running in the User Plane (UP). The block which provides the means for NFs to discover each other and related services is called the Network Repository Function (NRF).

\subsection{5G Network Slicing}
5G network slicing is a network architecture that allows for the creation of multiple virtualized and independent logical networks on the same physical network infrastructure. Each network slice is an isolated end-to-end network tailored to meet the diverse requirements of a specific application. This technology plays a central role in supporting 5G mobile networks, which are designed to efficiently accommodate a wide range of services with varying service level requirements. A network slice instance, as shown in Fig. \ref{5gc-arch}, is an end-to-end logical network (consisting of NF instances) custom defined to satisfy required networking characteristics and provide specific services to serve particular use cases (e.g., voice communication, video streaming, e-health, vehicular communication).


\section{Malicious Lateral Movement in 5GC WIth Network Slicing}
\label{attack-overview}

\subsection{Threat Model}
We assume that the attackers have compromised one of the NF containers in a given 5GC through one of many techniques such as exploiting NF interfaces exposed to Internet or installing their malicious containerized NF using an insider. Once the attackers have established a foothold in the 5GC, they need to move between NF containers and/or underlying hosts to reach their target slice/NF. It is also assumed that the attackers' movement is reflected in host/container logs. The attackers are expected to be well-resourced in terms of hacking capabilities, financial backing, computational resources and time as in the case of APT groups. 

\subsection{Lateral Movement Description}
NFs in 5GC are typically deployed as containers which is a virtualization technology to package together all resources needed to run an application on almost any type of platform.  Lateral movement in IT networks is an established attack technique wherein attackers move from their initial point of entry host in the network to other hosts or networks within an organization. Similarly, attackers can compromise a container/cluster and move laterally to other containers in the cluster.  A few techniques to gain unauthorized access to containers/cluster are exploiting misconfigured clusters, using stolen credentials/exploiting vulnerabilities to compromise a container/node, intercepting insecure traffic between nodes or nodes and the API server, exploiting Docker/Kubernetes components and container escape. 

Adversaries have been actively targeting real-world enterprise containers. Cloud-management platform Uptycs and cybersecurity services firm Crowdstrike have reported attacks on their honeypots (Docker servers), which were accessible through remote APIs, to install malicious container images for cryptomining and attacking important websites \cite{dark-reading}. Microsoft Defender for Cloud team has reported that the Kinsing malware was compromising Kubernetes clusters by exploiting known remote command execution (RCE) vulnerabilities in apps installed in container images and misconfigured containers running PostgreSQL database servers \cite{bleeping-comp}. As 5GC implementations become widespread, NF containers will be targeted by cyber attackers.

\begin{figure}[h]
	\centering
	\includegraphics[scale=0.35]{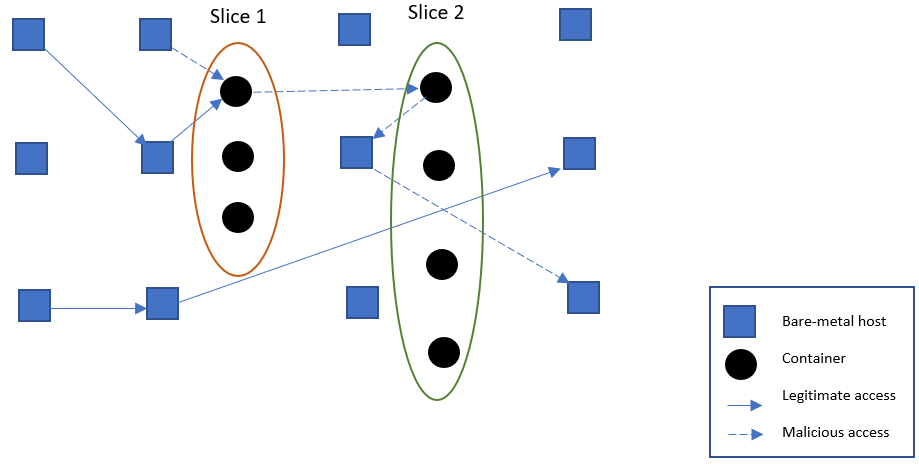}
	\caption{Lateral movement in a 5GC depicted as blue, dashed arrows}
	\label{5G-lat-mov-example}
\end{figure}

An attacker can mount mainly two types of LM (Fig. \ref{5G-lat-mov-example}) in a 5G network slicing deployment such as the one shown in Fig. \ref{5g-test-env-arch}. We have illustrated the LM types through examples below:
\begin{itemize}
	\item \textit{Same-slice LM}- The attacker gains access to the UPF container in slice 3 through one or more of the techniques to gain unauthorized access to containers. From there, the attacker can escape to the underlying host and compromise other containers in Slice 3 such as SMF and NRF.
	\item \textit{Cross-slice LM}- If the attacker gains access to the UPF container in Slice 1 through one or more of the unauthorized container access techniques, they can escape to the underlying host. Since the NRF container is shared between Slices 1 and 2, the attacker can pivot to Slice 2 (which it had no access to earlier).  From there, the attacker can compromise SMF and UPF containers in Slice 2.
\end{itemize}


\subsection{Lateral Movement Illustrative Scenarios}
\textit{Using remote access service}: An adversary in control of a compromised NF container may use a remote access service (e.g., SSH) to log in to another NF container using stolen credentials. Of course, this assumes that the remote access service is enabled on the target container, incoming connections are not blocked by the target container firewall and no other security hardening measures are in place on the target container (e.g., access to selected users only, password-based login disabling). In the context of 5GC implemented by OAI, NF Docker containers do not have any remote access service enabled by default. However, in production environments, certain NF containers have remote access enabled for maintenance \cite{trend-micro-5gc-vuln}. Further, even if we assume that system hardening is in place, there are bound to be mis-configurations (e.g., in firewall settings) and weakening of hardening compliance posture (e.g., due to deployment of new NFs or removal of existing ones) in at least some private 5GCs due to the complex environment with multiple software components.

\textit{Using container escape}: An adversary in control of a compromised NF container may use container escape techniques (e.g., shell code injection, process debugging) to escape to the underlying host and and from there gain access to other NF containers deployed on the same host. There are several container escape techniques, most of which use capabilities, which are a way to assign specific privileges to a running process. Capabilities can be applied to container processes; in this way, all the processes which are part of that container can inherit its capabilities. When a capability is assigned to a container, the caller thread can launch a set of system calls associated with the capability. A few examples of capabilities are SYS\_MODULE and SYS\_ADMIN. In order for an adversary to escape from a compromised NF container to the underlying host, the container should have certain capabilities enabled, and the specific container escape technique used depends on those capabilities. Further, the process ID (PID) address space between the NF container and the host OS should be shared. 

\textit{Using container misconfiguration}: An adversary in control of a compromised NF container/host may exploit container misconfiguration on a target host to gain privileges over the underlying host system and thus gain access to other NF containers deployed on that host. For example, the container service on the target host may be exposed through a TCP port with unencrypted communication enabled. The adversary can make a remote request through the container service API from the compromised host to access the container service running on the target host. On the successful grant of request, the adversary would be able to access all the NF containers deployed on target host. 	


\section{5GLatte Design}


%

\begin{figure}[h]
	\centering
	\includegraphics[scale=0.25]{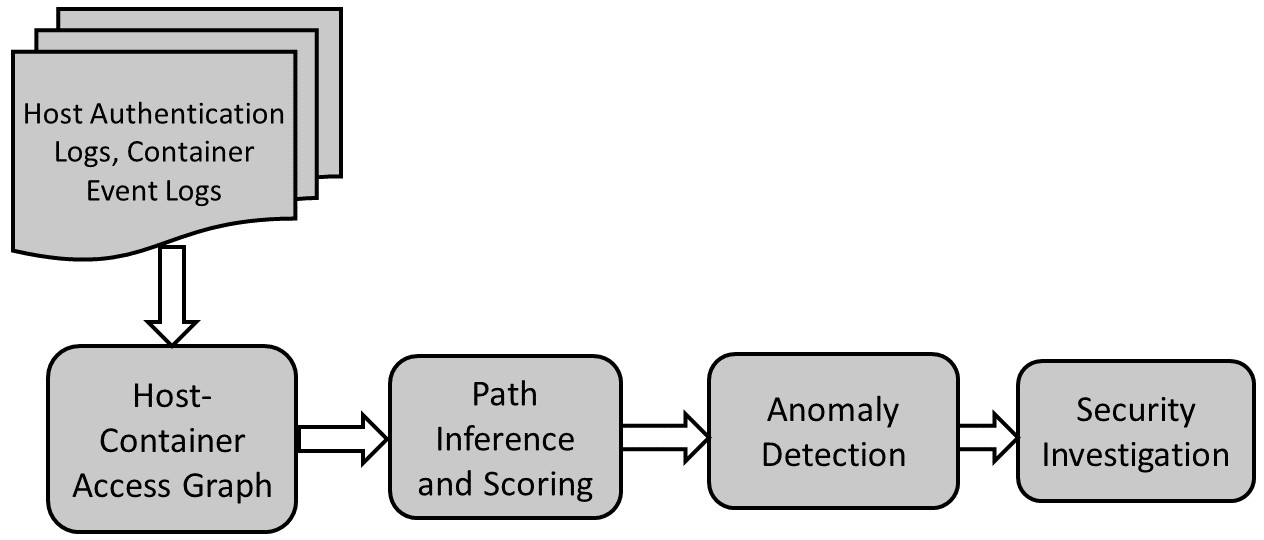}
	\caption{\textsf{5GLatte} Architecture}
	\label{5glatte-arch}
\end{figure}

The architecture of \textsf{5GLatte} is shown in Fig. \ref{5glatte-arch}. The first stage consists of building a host-container access graph using host/container logs collected from the 5GC. The next stage consists of inferring paths from the access graph and scoring them based on selected filtering criteria. Subsequently, the scored paths are presented as input to a threshold-based anomaly detection algorithm to reveal malicious lateral movement paths which can be forwarded to cyber security analysts for further investigation and action.

\subsection{Host-Container Access Graph Construction}
\begin{figure}[h]
	\centering
	\censorbox{%
	\includegraphics[scale=0.3]{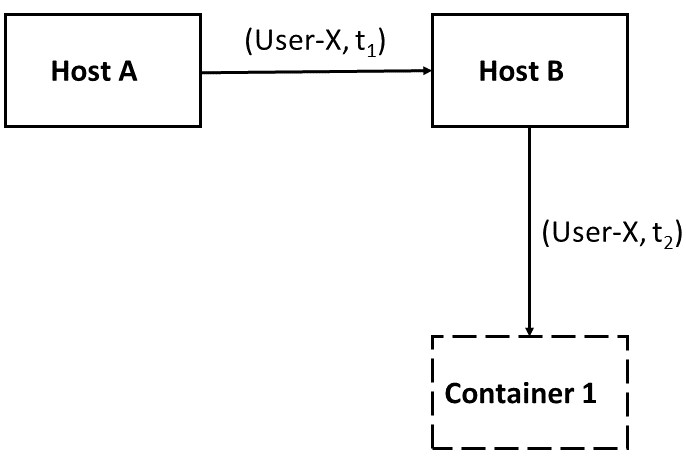}%
	}
	\caption{Example of a simple host-container access graph in a 5GC}
	\label{5G-host-cont-graph}
\end{figure}

\blackout{We construct a directed graph which captures the interactions between users and hosts/containers in a 5GC using host authentication logs and container event logs. The graph's edges represent login to a host from another connected host or command execution on a container from the underlying host or escape from a container to the underlying host. The source and destination nodes for an edge represent the host/container initiating the access and the host/container receiving the access, respectively. Each edge is embedded with the time-stamp of access and the user account being accessed on the destination node. These graphs serve as input to the later steps of the proposed LM detection system. A visualization of the graph is shown in} Fig. \ref{5G-host-cont-graph}. \blackout{The host authentication logs (e.g., Linux login logs, Windows Kerberos logs) capture authentication activity on a host in terms of the source machine name/IP address, the user account which is used to log in to the host, the date/time of login and logout and the service accessed. Container event logs capture various events (e.g., container creation, start, stop) on containers managed by a server or belonging to a cluster in terms of the date/time of the event, `id' of the container where the event has occurred and the type of event. These include events when a container is accessed from the host or remotely through the API server for example, using an interactive shell.}


\subsection{Path Score Computation}
\label{path-score}
\blackout{We define a path inferred from the host-container access graph as a series of connected edges, all of which are caused by the same user, with consecutive edges appearing within a certain maximum time threshold. The cumulative score for a path is calculated as a product of four sub-scores. The first path sub-score is a function of the non-zero minimum of historical edge probabilities for all the host-to-host edges constituting that path. Here, historical edge probability is the fraction of days in the past training data (collected for certain number of days) that a particular edge between two nodes in the access graph has been part of a successful login/access). The second path sub-score is a function of the non-zero minimum of historical edge probabilities for all the host-NF container edges which are part of that path. The third path sub-score is a function of the number of network slices that the path passes through. The fourth and final path sub-score is a function of the number of hops in that path.}


\subsection{Anomaly Detection} 
\blackout{To detect if a new path, $p$ is anomalous and is part of LM in an attack campaign, we calculate the cumulative score for that path as outlined in the previous sub-section and compare it with a threshold ($\alpha$). The threshold is calculated as the maximum cumulative path score generated during training, i.e, when the given 5GC is not under attack. The anomaly detection can be expressed mathematically as:}



\section{Implementation}
\begin{figure*}[t]
	\caption{5G Network Slicing Deployment}
	\label{5g-test-env-arch}
	\centering
	\includegraphics[scale=0.3]{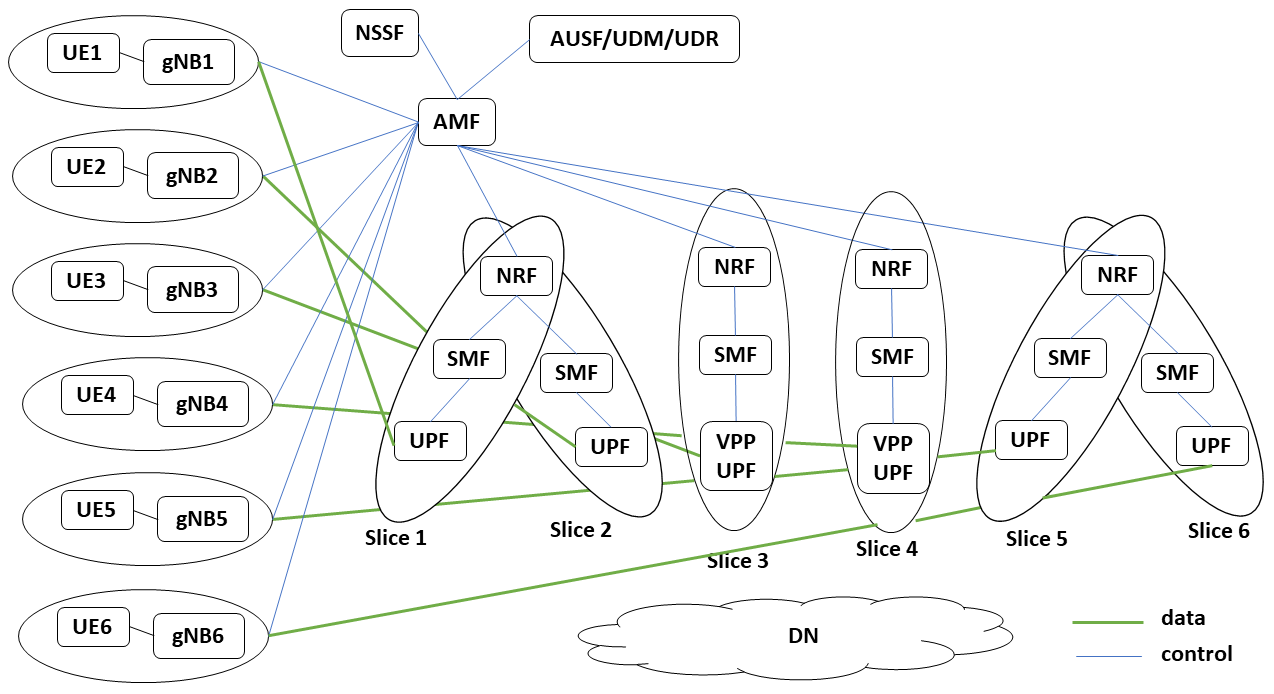}
\end{figure*}

To evaluate the performance of \textsf{5GLatte}, we built a 5G test environment consisting of network slices using Open Air Interface (OAI) software which includes Docker containers corresponding to the various 5G core network components (e.g., AMF, AUSF, NSSF, NRF, SMF, UPF) as well as the OAI gNB and NR-UE emulators. The test environment architecture is shown in Fig. \ref{5g-test-env-arch}. The Docker container images were deployed on a VMWare ESXi server VMs with Intel Xeon Silver 4216 CPU @2.10GHz, 64-bit architecture, 8 cores, 16GB RAM and running Ubuntu 18.04/Ubuntu 20.04 OS. We also tested Internet connectivity from the NR-UEs successfully. \textsf{5GLatte} was implemented in Python 3.6 with the \textit{networkx} library being used to construct host-container graphs. Our implementation was not designed for real-time operation.

\subsection{Experimental Methodology}
\textit{Training phase}: We ran the 5G test environment without any attacks on day 1 and collected data from it. This served as training data for \textsf{5GLatte}. During this phase, it is assumed that the MNO remotely accesses the UPF in all network slices from Internet and locally accesses the other NFs from their underlying hosts for maintenance purposes. This is a feasible scenario since the UPF interface has been found to be exposed to the Internet in certain topologies followed in private 5G network deployments \cite{trend-micro-5gc-vuln}. \blackout{The threshold with which we compare the anomaly score for an LM path during the testing phase as part of the anomaly detection algorithm explained in Section path-score was determined using this training data.} 

\textit{Testing phase}: We ran multiple attack campaigns in the 5G test environment (while also performing normal operations such as accessing NFs for maintenance) and collected data from it. Specifically, we ran \textit{Attack Campaign 1} with sub-scenarios \#1 and \#2 on days 2 and 3 respectively, \textit{Attack Campaign 2} on day 4 and \textit{Attack Campaign 3} with sub-scenarios \#1 and \#2 on days 5 and 6 respectively. The sub-scenarios \#1 and \#2 are mutually exclusive, i.e., at one time only one of them are used as part of an attack campaign. The data so collected served as test data for evaluating the performance of \textsf{5GLatte}. In all the above attack campaigns, we assume that the adversaries belong to an APT group which has been tasked with targeting a critical 5G network slice, e.g., a government e-health network to access confidential data and/or disrupt the provision of health services to legitimate customers. Specifically, the adversaries aim to drop the health services customers’ data traffic by gaining control of the UPF in e-health network slice. The steps followed in each attack campaign are based on the attack stages in MITRE ATT\&CK (containers) and FiGHT frameworks, and have been outlined below.  

\textbf{Attack Campaign 1}:
\begin{enumerate}
	\item The attacker searches for Docker/Kubernetes API servers with exposed ports and finds one to gain access to AMF.
	\item The attacker downloads and runs a malware executable to run a reverse shell back to an attacker-controlled server from the compromised container.
	\item The attacker runs a scan for other containers in the network and finds NRF, AUSF and NSSF.
	\item The attacker escapes to the underlying host using process debugging technique.
	
	\noindent\textbf{Sub-scenario \#1}
	\begin{enumerate}
		\item Attacker runs a scan for other hosts and finds one running docker daemon but with closed ports.
		\item It scans internal file system of the host and finds password hashes. It uses them to login to another host in the network.
		\item The other host deploys containers in slice 3 (NRF, SMF, UPF). The attacker uses \textit{exec} to run commands on those containers.
		\item The attacker queries the NRF for slice identifier (SD) to confirm that it is the target slice.
	\end{enumerate}
	
		\noindent\textbf{Sub-scenario \#2}
	\begin{enumerate}
		\item Attacker runs a scan for other hosts and finds one running \textit{docker} daemon but with closed ports.
		\item It scans internal file system of the host and finds password hashes. It uses them to login to another host in the network.
		\item The attacker uses \textit{exec} to run commands on the containers deployed on the other host.
		\item The attacker queries the NRF for slice identifier (SD) to confirm if it is the target slice.
		\item On finding that it is not the target slice, the attacker repeats the previous four steps until it reaches the host deploying target slice (slice 6) containers.		
	\end{enumerate}
	\item The attacker runs a script on those containers and establishes a connection back to C\&C server.
	\item The attacker accesses the UPF and redirects the users’ traffic, causing DoS. 

\end{enumerate}

\textbf{Attack Campaign 2}:
\begin{enumerate}
	\item The attacker searches for Docker/Kubernetes API servers with exposed ports and finds one to gain access to UPF in slice 5.
	\item The attacker downloads and runs a malware executable to run a reverse shell back to an attacker-controlled server from the compromised container.
	\item The attacker runs a scan for other containers in the network and finds NRF and SMF. The attacker looks at the container configuration and find that it is part of a 5G network slice with a certain SD. The attacker queries the NRF for slice identifier (SD) to confirm if it’s the target slice. It finds out that the current slice (slice 5) is not the target slice.
	\item The attacker escapes to the underlying host using process debugging technique.
	\item The attacker lists VLANs on the host and finds two of them corresponding to slice 5 and 6. The attacker finds the NF containers which are part of VLAN 2 (slice 6). It also checks the container configuration to find the slice 6 SD which is target slice. 
	\item The attacker runs commands on the slice 6 containers.
	\item The attacker runs a script on those containers and establishes a connection back to C\&C server.
	\item The attacker accesses the UPF in slice 6 and redirects the users’ traffic, causing DoS.  
	
\end{enumerate}

\textbf{Attack Campaign 3}:
\begin{enumerate}
	\item The attacker searches for Docker/Kubernetes API servers with exposed ports and finds one to gain access to UPF in slice 1.
	\item The attacker downloads and runs a malware executable to run a reverse shell back to an attacker-controlled server from the compromised container.
	\item The attacker runs a scan for other containers in the network and finds NRF and SMF. The attacker looks at the container configuration and find that it is part of a 5G network slice with a certain SD. The attacker queries the NRF for slice identifier (SD) to confirm if it’s the target slice. It finds out that the current slice (slice 1) is not the target slice.
	\item The attacker escapes to the underlying host using process debugging technique.

	\noindent\textbf{Sub-scenario \#1}
	\begin{enumerate}
		\item Attacker runs a scan for other hosts and finds one running docker daemon but with closed ports.
		\item It scans internal file system of the host and finds password hashes. It uses them to login to another host in the network.
		\item The other host deploys containers in slice 3 (NRF, SMF, UPF). The attacker uses \textit{exec} to run commands on those containers.
		\item The attacker queries the NRF for slice identifier (SD) to confirm that it is the target slice.
	\end{enumerate}
	
		\noindent\textbf{Sub-scenario \#2}
	\begin{enumerate}
		\item Attacker runs a scan for other hosts and finds one running \textit{docker} daemon but with closed ports.
		\item It scans internal file system of the host and finds password hashes. It uses them to login to another host in the network.
		\item The attacker uses \textit{exec} to run commands on the containers deployed on the other host.
		\item The attacker queries the NRF for slice identifier (SD) to confirm if it is the target slice.
		\item On finding that it is not the target slice, the attacker repeats the previous four steps until it compromises the host deploying slice 4 containers.		
	\end{enumerate} 
	\item From the slice 4 host, the attacker makes a remote request through the container API to the connected host, which is successfully granted. Using the container service runtime, the attacker checks the container configuration to find the slice 6 SD and then queries it with the NRF to confirm that it is the target slice.
	\item It is able to access all the NF containers deployed on the slice 6 host using container service runtime.
	\item The attacker runs a script on those containers and establishes a connection back to C\&C server.
	\item The attacker accesses the UPF in slice 6 and redirects the users’ traffic, causing DoS.
	
\end{enumerate}

\subsection{Performance Evaluation Results}
In Table \ref{tpr-fpr-results}, we have shown the detection performance results of \textsf{5GLatte} on our small-scale 5G attack dataset in terms of TP (True Positives), FP (False Positives), TPR (True Positive Rate) and FPR (False Positive Rate). TPR is the ratio $\frac{TP}{TP+FN}$, where $FN$ is the number of false negatives, FPR is the  ratio $\frac{FP}{FP+TN}$, where $TN$ is the number of true negatives. \blackout{Here, a positive sample refers to an edge in the host-container access graph which is part of the lateral movement while a negative sample refers to an edge in the host-container access graph which is not a part of the lateral movement.} \textsf{5GLatte} achieved a TPR for 100\% with an FPR of 8.33\%. Though \textsf{5GLatte} performs well in terms of TPR, the FPR is on the higher side as well. However, such high FPR might be due to the small-scale simulated test environment. Even a single false positive increases the FPR significantly. We have compared the detection performance of \textsf{5GLatte} with two LM detectors based on Hopper \cite{hopper} \blackout{which use only the first two path sub-scores and the first three path sub-scores mentioned in sub-section path-score, respectively}. The \textit{Hopper-1} detector performs worst, with a TPR of 0\% and an FPR of 8.33\%. The \textit{Hopper-2} detector performs better than \textit{Hopper-1} but worse than \textsf{5GLatte}, with a TPR of 84.21\% and an FPR of 8.33\%. 

\begin{table}[h]
	\centering
	\begin{tabular}{|l|l|l|l|l|l|l|}
	\hline
	\textbf{Algorithm} & \textbf{TP} & \textbf{FP} & \textbf{TPR (\%)} & \textbf{FPR (\%)}  \\ \hline
	5GLatte & 19 & 1 & 100 & 8.33 \\ \hline
	Hopper-1 & 0 & 1 & 0 & 8.33 \\ \hline
	Hopper-2 & 16 & 1 & 84.21 & 8.33 \\ \hline
	\end{tabular}
	\caption{\textsf{5GLatte} Detection Performance Results}
	\label{tpr-fpr-results}
\end{table}





\subsection{Discussion}
\subsubsection{False Positives}: \blackout{It should be noted that 5GLatte is not a generalized threshold-based anomaly detection technique. We are looking for a specific attack signature in terms of the path features listed in Section path-score which help us to filter malicious LM paths. However, it is possible that some legitimate path instances might also satisfy those features which leads to false positives. Since false positives can lead to additional overhead for cyber security analysts investigating threats, they must be minimized as much as possible. This can be achieved by designing additional filters to reduce false positives. One such filter can be to discard any LM paths detected which involve only one host-NF container edge which has been observed earlier in the training data. In our case, using this filter helps bring down the number of false positives for 5GLatte to zero. However, this might be because we have used a small-scale dataset. For a larger dataset, deploying the filter might reduce the number of false positives but may not bring it down to zero.}

\subsubsection{False Negatives}: \blackout{We did not observe any false negatives in the performance evaluation of 5GLatte with our small-scale dataset. It is possible for 5GLatte to produce false negatives with a larger dataset and in real-world settings, false negatives would need to be minimized as well to reduce threat investigation overhead. However, the direct connection between NF containers and underlying hosts in the 5GC architecture and the infrequent need to access NF containers in network slices once deployed mean that there is a low likelihood of 5GLatte mistaking a malicious LM path for a legitimate NF container access unless the adversary is actively trying to mask the LM path with legitimate container accesses.}  

\subsubsection{Limitations}: \blackout{Since 5GLatte uses a set of path features as listed in Section path-score for filtering malicious LM paths, an adversary aware of those features may follow a LM path that is designed to evade detection. For example, the attacker's path may chain together multiple legitimate NF container accesses. However, this may not always be possible depending on the location of the target network slice/NF container. Another weak link is the host authentication logs and container event logs used to construct the host-container access graph. If those logs are not present or do not contain accurate information, it may result in false negatives. An adversary aware of 5GLatte's design may also delete those logs on purpose if they are not read-only, thus subverting the detection process. Finally, at nodes along their path to the target slice/container, the adversary could (1) wait until the maximum time threshold used by 5GLatte for linking consecutive host-container graph edges has elapsed and then continue to access the next host/container, or (2) change to a different user account to evade detection. However, waiting delays the adversary's advance through the 5GC, increasing the risk of detection by container protection tools and host file system scanners while changing to a different user account depends on the availability of such accounts and the adversary's system privileges.}




\section{Conclusion}
In this paper, we have proposed malicious lateral movement strategies in a 5G Core with network slicing enabled. We have also designed \textsf{5GLatte}, a system to detect such lateral movement which uses threshold-based anomaly detection on paths inferred from host-container access graphs built with host/NF container logs collected from the 5GC. \textsf{5GLatte} detects malicious lateral movement with a higher true positive rate and similar false positive rate compared to other lateral movement detectors based on state-of-the art on a dataset containing attack campaigns (based on MITRE ATT\&CK and FiGHT frameworks) launched in a 5G test environment. 


\bibliographystyle{ieeetran}
\begingroup
\raggedright
\bibliography{5glatte}
\endgroup

%

\end{document}